\documentstyle[prl,aps,floats]{revtex}

\begin{document}
\preprint{PU-RCG-00/5, hep-th/0002215}
\draft

%
%
\input epsf

\twocolumn[\hsize\textwidth\columnwidth\hsize\csname
@twocolumnfalse\endcsname

\title{Cosmological perturbation spectra from
SL(4,R)-invariant effective actions}
\author{Helen A. Bridgman and David Wands}
\address{Relativity and Cosmology Group, School of Computer Science
and Mathematics, Mercantile House, University of Portsmouth,
Portsmouth PO1 2EG, United Kingdom}
\maketitle
\begin{abstract}
We investigate four-dimensional cosmological vacuum solutions derived
from an effective action invariant under global SL($n$,R)
transformations. We find the general solutions for linear axion field
perturbations about homogeneous dilaton-moduli-vacuum solutions for an
SL($4$,R)-invariant action and find the spectrum of super-horizon
perturbations resulting from vacuum fluctuations in a pre-big-bang
scenario. We show that for SL($n$,R)-invariant actions with $n\geq4$
there exists a regime of parameter space of non-zero measure where all
the axion field spectra have positive spectral tilt, as required if
light axion fields are to provide a seed for anisotropies in the
microwave background and large-scale structure in the universe.
\end{abstract}

\pacs{PACS numbers: 98.80.Cq \hfill PU-RCG-00/5, hep-th/0002215}
\pacs{to appear in Physical Review D \hfill \copyright 2000 The American Physical Society}
\vskip2pc]

\section{Introduction}

The pre-big-bang scenario proposed by Gasperini and Veneziano
\cite{pbb} is an alternative model for the very early evolution of our
Universe which assumes that its initial state was a low-energy, weakly
coupled, vacuum state. Such a regime is well described by the
low-energy string effective action which admits two separate branches,
labelled $(+)$ and $(-)$, for vacuum solutions of the scale factor in
four-dimensional Friedmann-Robertson-Walker (FRW)
cosmologies\cite{gracefulexit}. The $(+)$ branch corresponds to a
weakly coupled dilaton in a cold, flat universe in the $t \to -\infty$
limit.  For $t \to 0$ we have pole driven super-inflation propelled by
the dilaton kinetic energy term, with a positive Hubble parameter, $H$
and a singularity in the future. The $(-)$ branch corresponds to a
large spatially flat universe with positive $H$, but decelerating and
can be smoothly joined to a conventional radiation dominated universe
at late times.  The $(+)$ and $(-)$ branches are related to each other
by a string symmetry called scale factor duality~\cite{MeiVen91,pbb},
but there is still no compelling dynamical model of the ``graceful
exit" from the $(+)$ to the $(-)$ branch \cite{gracefulexit,BruMad97}.

In the absence of a complete theoretical understanding one may still
hope to find observational evidence, such as the spectrum of
primordial fluctuations that could be generated during the
dilaton-driven pre-big-bang phase~\cite{LidWanCop99}, in order to test
the scenario against more conventional inflation models. Metric
perturbations are produced on super-horizon scales during the pre big
bang have a steep ``blue'' spectrum, strongly tilted towards small
scales~\cite{BruGasGio95}. This offers the interesting possibility
that there might be a detectable background of relic gravitons on
Laser Interferometric Gravitational wave Observatory (LIGO)
scales~\cite{AllBru97,UngVec99} and a related population of primordial black
holes~\cite{CopLidLid98}. However these metric perturbations are far
from the almost scale-invariant (Harrison-Zel'dovich) spectrum of
adiabatic density perturbations naturally produced by conventional
slow-roll inflation models~\cite{LidLyt93} and leave effectively no
metric perturbations on astrophysical scales during the pre-big-bang
era.

Instead it has been proposed~\cite{durrer98} that a cosmic background
of massless axion fluctuations could generate the observed
anisotropies in the cosmic microwave background temperature at large
angular scales and provide a seed for large-scale structure
formation~\cite{durrer98,durrer99,MelVerDur99}. A pre-big-bang era can
produce almost scale invariant spectra of fluctuations in axion
fields~\cite{vac fluctn,BruHad98,Buonanno}
$\langle\delta\sigma^2\rangle\propto k^{\Delta n}$ with $\Delta
n\approx0$, where $k$ is the comoving wavenumber. These are
isocurvature perturbations to first-order, but assuming the axion
field remains effectively massless in the subsequent
post big bang era, these fluctuations give rise to a spectrum of
density perturbations at horizon crossing~\cite{durrer98}
\begin{equation}
\label{one}
\left({\delta\rho\over \rho}\right)_{k=aH}
 \sim e^\phi \left( {k\over k_s} \right)^{\Delta n} \ ,
\end{equation}
where $k_s$ is the comoving scale leaving the horizon at the end of
the pre-big-bang phase. A slightly ``blue'' spectrum, $\Delta
n\approx+0.1$, may be consistent with $\delta\rho/\rho\sim 10^{-5}$ on
astrophysical scales ($k\sim10^{30}k_s$) for a present-day
string coupling $g_s^2=e^\phi\sim10^{-2}$.

The low energy string effective action compactified down to
four-dimensions includes a dilaton and axion field, related by an
SL(2,R) symmetry. In the absence of all the other moduli fields
are fixed during the pre-big-bang phase then it is found that for
an SL(2,R) action, with one axion field, the spectral index is
fixed to be $\Delta n=-2\sqrt{2}+3=-0.46$. The addition of a
single moduli field gives a spectral index for the axion in the
range $-0.46\leq \Delta n \leq 3$ which allows scale-invariant or
blue spectra~\cite{vac fluctn}.

The many moduli fields present in any low-energy effective action
will have specific symmetry properties inherited from the higher
dimensional theory and the details of compactification. For
instance, the inclusion of Ramond-Ramond (RR) fields presents in
the type II string theories increases the number of degrees of
freedom in the four-dimensional effective theory and in
Ref.~\cite{symm gen str cos} it was shown that the RR 1-form and
3-form field strengths, with a single modulus field determining
the size of the 6-torus, combine with the Neveu-Schwarz-Neveu-Schwarz(NS-NS) dilaton and axion
to parameterise an SL(3,R)-invariant non-linear sigma model. The
symmetries of this action can place constraints on the allowed
spectral indices. For an SL(3,R) action \cite{axion pertbn}, with
two moduli but three axion fields, the range for each spectral
index was the same as for the single axion case ($-0.46 \leq
\Delta n_i \leq 3,\ i=1,2,3$), but there was no point at which all
the spectra had $\Delta n_i>0$. This poses a threat to the pre-big-bang scenario as all the perturbation spectra have the same
normalisation at the string scale ($\delta\rho/\rho\sim e^\phi$)
and if one axion field always has a red spectrum ($\Delta n_i<0$)
then there would be unacceptably large density fluctuations on
large scales.

In order to study whether this remains a problem in larger
symmetry groups with more moduli and axion degrees of freedom we
will study the spectrum of cosmological perturbations generated in
fields which parameterise an SL(4,R) non-linear sigma model in the
low-energy effective action.  The presence of a global SL($n$,R)
symmetry is a completely general consequence of dimensional
reduction from $D+n$ to $D$
dimensions~\cite{CreJul98,LidWanCop99}.  We will investigate the
perturbation spectra generated in axion fields at late times or large
scales from vacuum fluctuations at early times or small scales in a
pre-big-bang era, and discuss whether these might be compatible
with an almost scale invariant spectrum of small primordial
density perturbations.  We will show that perturbation spectra
with $\Delta n>0$ for all fields are indeed possible in models
whose scalar fields parameterise an SL($n$,R) group where
$n\geq4$.

\section{SL($n$,R) invariant actions}

We begin with a discussion of the representation of the most
familiar SL(2,R)-invariant effective action is string theory, and
how we can extend this to provide a representation of larger
SL($n$,R)-invariant effective actions.

The NS-NS sector of string theory contains the dilaton, $\phi$, graviton, $\hat{G}_{AB}$, and 2-form
potential, $B_{AB}$, and is common to both heterotic and type II string
theories. The low energy effective action is
\begin{equation}
\label{10d}
S = {1\over 16\pi\alpha^{\prime4}}
\int d^{10}x \sqrt{|\hat{G}|} \, e^{-\phi} \left[ \hat{R}_{10} +
(\hat\nabla\phi)^2 - {1\over12} \hat{H}^2 \right] \,,
\end{equation}
where $\alpha'$ is the inverse string tension and $\hat{H}\equiv dB$
is a 3-form field strength.
Considering the simplest Kaluza-Klein compactification on a static
six-torus, the ten-dimensional line element has the form
$ds_{10}^2=e^{\phi}g_{\mu\nu}dx^{\mu}dx^{\nu}+\delta_{ab}dy^ady^b$,
where $g_{\mu\nu}$ is the four-dimensional metric in the Einstein
frame and $\delta_{ab}$ is the Kronecker delta-function.
In four-dimensions the 3-form field strength is dual to a one-form
written as the gradient of a pseudo-scalar axion field,
$*H=e^\phi\nabla\sigma$, and the four-dimensional effective action is
then~\cite{LidWanCop99}
\begin{equation}
\label{4dNSNS}
S = \frac{1}{2\kappa^2}\,\int d^4x\ \sqrt{-{g}}\
\left [ {R} -
\frac{1}{2}\,\left({\nabla}\phi \right)^{2} -
\frac{1}{2}\,e^{2\phi} \left({\nabla}\sigma
\right)^{2}\right] \,,
\end{equation}
where $\kappa^2 \equiv 8\pi \mathrm{G}$. Solutions to the
equations of motion from this action respect the invariance of
this action under an arbitrary global SL(2,R)
transformation~\cite{ShaTriWil91}
\begin{equation}
\label{eq:trans}\lambda \to \frac{\alpha \lambda + \beta}{\gamma
\lambda + \delta} \,,
\end{equation}
where $\lambda = \sigma + i e^{-\phi}$ and the real parameters
$\alpha$, $\beta$, $\gamma$ and $\delta$ obey the constraint $\alpha
\delta - \beta \gamma = 1$.

In order to extend the effective action to larger symmetry groups it
is convenient to re-write the action Eq. ~(\ref{4dNSNS}) in terms of the
symmetric matrix
\begin{equation}
\label{eq:sl2matrix}
M=
\left( \begin{array}{ll}
e^{\phi} & \sigma e^{\phi}\\
\sigma e^{\phi} & e^{-\phi}+\sigma^{2}e^{\phi}\\
\end{array}  \right)
\end{equation}
which parameterises the SL(2,R)/U(1) maximal coset of SL(2,R). The
SL(2,R) transformation in Eq. (\ref{eq:trans}) is given by $M \to
\Theta M \Theta^T$ where
\begin{displaymath}
\Theta = \left(\begin{array}{cc}
\delta & \gamma \\
\beta & \alpha \\
\end{array}\right) \,.
\end{displaymath}
Any member of SL(2,R) obeys the relation $M^TJM = J$ where
\begin{displaymath}
J= \left( \begin{array}{cc}
0 & 1 \\
-1 & 0 \\
\end{array} \right)
\end{displaymath}
and hence, using the expression $\Theta^T J \Theta = J$ we can show that
the line element
\begin{equation}
dS^2 = \frac{1}{2}{\mathrm{Tr}}(J dM J dM) = d\phi^2 +
e^{2\phi}d\sigma^2
\end{equation}
is invariant under the global SL(2,R) transformation $M \to \Theta M
\Theta^T$.
Thus the effective action Eq. ~(\ref{4dNSNS}) written in the form
\begin{equation}
S={\frac{1}{2\kappa^2}}\,\int {\mathrm{d^4}}x\ \sqrt{-g}\ \left [
{R} + \frac{1}{4}\:{\mathrm{Tr}} \left( {\nabla}M {\nabla}M^{-1}
\right)\right]
\end{equation}
is manifestly invariant under global SL(2,R) transformations.
More generally the action for scalar fields parameterising an
SL($n$,R)/SO($n$) non-linear sigma model can be written as
\begin{equation}
\label{slnaction}
S={\frac{1}{2\kappa^2}}\,\int {\mathrm{d^4}}x\
 \sqrt{-{g}}\ \left [ {R}
+ {1\over4} {\rm Tr} \left( \nabla U_n \nabla U_n^{-1} \right)
\right] \,,
\end{equation}
where $U_n$ is a symmetric SL($n$,R) matrix. This action is
invariant under the global transformation $U_n\to\Theta U_n
\Theta^T$ where $\Theta$ is a member of SL($n$,R).

We can build a symmetric SL(3,R) matrix $U$, from the SL(2,R)
matrix $M$ in the following way \cite{gal'tsov,symm gen str cos}:
\begin{equation}
\label{sl3matrix}
U =
\left( \begin{array}{ll}
e^{\nu} M & e^{\nu} M \sigma\\
e^{\nu} \sigma^{T}M & e^{-2\nu}+e^{\nu}\sigma^{T}M\sigma\\
\end{array}  \right) \,,
\end{equation}
where $\nu$ is a modulus field and the two additional degrees of
freedom are the components of the $2\times1$ vector
\begin{equation}
\mathbf{\sigma}=
\left( \begin{array}{c}
\sigma_2\\
\sigma_3\\
\end{array}
\right) \,.
\end{equation}
The SL(3,R)-invariant trace of the $3\times3$ matrix ${\nabla}U
{\nabla}U^{-1}$ which appears in the effective action is
\begin{eqnarray}
{\mathrm{Tr}} \left({\nabla}U {\nabla}U^{-1} \right) &=&
{\mathrm{Tr}} \left({\nabla}M{\nabla}M^{-1}\right) \nonumber \\
&& - 6\left({\nabla}\nu\right)^{2}
 - 2{\mathrm{Tr}}
\left(e^{3\nu}{\nabla}{\sigma}^{T}M{\nabla}{\sigma}\right) \,.
\end{eqnarray}

More generally, the same method can be used to construct an
SL($n+1$,R) matrix $U_{n+1}$ from an SL($n$,R) matrix $U_n$, where
\begin{equation}
\label{sln+1matrix}
U_{n+1}=
\left( \begin{array}{ll}
e^{v_{n+1}} U_n & e^{v_{n+1}} U_n \sigma_n\\
e^{v_{n+1}} {\sigma_n}^{T} U_n &
e^{-nv_{n+1}}+e^{v_{n+1}}{\sigma_n}^{T}U_n\sigma_n\\
\end{array}  \right) \,.
\end{equation}
In addition to the fields in the SL($n$,R) matrix $U_n$, the
SL($n+1$,R) matrix $U_{n+1}$ includes an additional modulus field
$v_{n+1}$ and $n$ additional axion fields contained in the
$n\times1$ vector $\sigma_n$. This is sufficient to define our
representation of the SL($n$,R)/SO($n$) coset starting from
$U_1=1$. Thus the SL($n$,R) matrix $U_n$ contains $n-1$ moduli and
$n(n-1)/2$ axions in total.

The SL(n+1,R)-invariant trace of the $(n+1)\times(n+1)$ matrix
${\nabla}U_{n+1} {\nabla}{U_{n+1}}^{-1}$ which appears in the
effective action [see Eq.~(\ref{slnaction})] can be calculated
iteratively as
\begin{eqnarray}
\label{eq:trace}{\mathrm{Tr}} \left({\nabla}U_{n+1}
{\nabla}{U_{n+1}}^{-1}
\right) &=&
{\mathrm{Tr}} \left({\nabla}U_n{\nabla}{U_n}^{-1}\right)
\nonumber\\&&
 + (-n^2-n)\left({\nabla}v_{n+1}\right)^{2}
\nonumber\\&&
\hspace{-24pt}
-2{\mathrm{Tr}}\left(
e^{(n+1)v_{n+1}}{\nabla}{\sigma}_n^{T}U_n{\nabla}{\sigma}_n
\right) \,.
\end{eqnarray}


\section{SL(4,R) Dilaton-Moduli-Vacuum Cosmologies}

We first investigate the homogeneous dilaton-moduli vacuum
solutions where we set $\sigma_i=$constant. The form of the
dilaton-moduli solutions are invariant under a constant shift of
the axion fields, so without loss of generality we set
$\sigma_i=0$. This amounts to setting to zero the off-diagonal
terms in the SL($n$,R) matrix. Thus we have the vacuum SL(2,R)
matrix (putting $\phi=v_2$)
\begin{equation}
\label{eq:sl2vac}
U{_2}^{(0)}=
\left( \begin{array}{cc}
e^{v_2} &0\\
0 & e^{-v_2}\\
\end{array}  \right)
\end{equation}
and the SL(3,R) matrix given in Eq. (\ref{sl3matrix}) becomes
\begin{equation}
\label{eq:sl3vac}
U{_3}^{(0)}=
\left( \begin{array}{ccc}
e^{v_3+v_2} &0 &0\\
0 & e^{v_3-v_2} &0\\
0 &0 & e^{-2v_3} \\
\end{array}  \right)
\end{equation}
while, from Eq.~(\ref{sln+1matrix}), an SL(n+1,R) matrix is given
in terms of an SL($n$,R) matrix as
\begin{equation}
U_{n+1}^{(0)}=
\left( \begin{array}{ll}
e^{v_{n+1}} U_n^{(0)} & 0\\
0 & e^{-nv_{n+1}} \\
\end{array}  \right) \,.
\end{equation}

The trace of ${\nabla}U_n^{(0)}{\nabla}{U_n^{(0)}}^{-1}$ for a
vacuum SL($n$,R) matrix, using Eq. (\ref{eq:trace}), is
\begin{equation}
\label{slnvac} {\rm
Tr}\left({\nabla}U_n^{(0)}{\nabla}{U_n^{(0)}}^{-1}\right) = -
\sum_i^n i(i-1) \left( \nabla v_i \right)^2 \,.
\end{equation}
Substituting this into the SL($n$,R)-invariant action,
Eq.~(\ref{slnaction}), we can now write down the effective
action for vacuum SL(4,R)
\begin{eqnarray}
\label{sl4vacaction} S&=&\frac{1}{2\kappa^2}\,\int
{\mathrm{d^4}}x\ \sqrt{-{g}}\ \nonumber \\ && \times \, \left[\ {R}
-3({\nabla}v_4)^2
-\frac{3}{2}\,({\nabla}v_3)^2-\frac{1}{2}\,({\nabla}v_2)^2\right]
\,.
\end{eqnarray}


We will assume that the external four dimensional spacetime is a
spatially-flat FRW metric, with the line element
\begin{equation}
ds^2= a^2(\eta)\left(-\mathrm{d}\eta^2 + \mathrm{d}x^2 +
\mathrm{d}y^2 + \mathrm{d}z^2 \right)
\end{equation}
and scale factor, $a(\eta)$, where $\eta$ is the conformal time.
%
The effective action in Eq.~(\ref{sl4vacaction}) can then be written
for homogeneous dilaton-moduli fields as
\begin{eqnarray}
\label{eq:action} S&=&\frac{1}{2\kappa^2}\,\int
{\mathrm{d^3}}x\int {\mathrm{d}}\eta\ \nonumber \\ && \qquad
\left[ 6{a'}^2
-3a^2{{v_4}'}^2-\frac{3}{2}\,a^2{{v_3}'}^2-\frac{1}{2}\,a^2{{v_2}'}^2\right]
\,,
\end{eqnarray}
where a prime denotes differentiation with respect to $\eta$.
We use the Euler-Lagrange equations derived from this action to
calculate the equations of motion related to this action.
The resulting evolution equations for the dilaton-moduli
fields in a FRW metric are
\begin{eqnarray}
\label{eq:a2prime}
12\frac{a''}{a}&=&-6{{v_4}'}\,^2-3{{v_3}'}\,^2-{{v_2}'}\,^2 \,,\\
\label{eq:v4} {v_4}''+2\frac{a'}{a}\,{v_4}'&=&0 \,,\\
\label{eq:v3} {v_3}''+2\frac{a'}{a}\,{v_3}'&=&0 \,,\\
\label{eq:v2} {v_2}''+\frac{a'}{a}\,{v_2}'&=&0 \,,
\end{eqnarray}
subject to the constraint equation (from general
relativistic invariance under time-reparameterisation)
\begin{equation}
\label{eq:constraint}
\left( {a'\over a} \right)^2 =
\frac{1}{12}\,\left(6{{v_4}'}^2+3{{v_3}'}^2+{{v_2}'}^2\right) \,.
\end{equation}
%

Substituting Eq. ~(\ref{eq:constraint}) into
Eq. ~(\ref{eq:a2prime}) yields a second-order equation for
$a(\eta)$ which can be integrated twice to obtain
\begin{equation}
\label{eq:a} a=a_*|\eta|^{\frac{1}{2}} \,,
\end{equation}
where $a_*$ is one integration constant, and we have used our
freedom to choose $a=0$ as the origin for the time coordinate
$\eta$ in order to eliminate the other constant of integration.
Equation~(\ref{eq:a}) is the standard solution in the Einstein
frame for a spatially flat FRW cosmology with free scalar fields.
In terms of the proper time $t\equiv\int ad\eta$ we have $a\propto
|t|^{1/3}$, which describes a non-accelerating expanding universe
for $t>0$, or an accelerating contraction for $t<0$. It is this
phase of accelerated contraction that is the basis of the pre-big-bang scenario~\cite{pbb}.

Equations (\ref{eq:v4}) to (\ref{eq:v2}) can now be integrated using
the solution for $a(\eta)$ in Eq.~(\ref{eq:a}) to give
\begin{equation}
\label{eq:consts} {v_i'}=\frac{C_i}{a^2} \,,
\end{equation}
where $C_i$ are constants of integration, $i=1 \ldots 3$.
This allows the constraint equation~(\ref{eq:constraint}) to be
rewritten in the form
\begin{equation}
{C_1^2 \over 3} + {C_2^2} + {2C_3^2} = 1 \,.
\end{equation}
The constants of integration
$C_1/\sqrt{3}$, $C_2$ and $\sqrt{2}C_3$ can be
interpreted as points on the surface of a sphere.
It is therefore convenient to move to spherical coordinates
\begin{eqnarray}
C_1/\sqrt{3} &=& \cos{\xi_1}  \,, \nonumber\\ C_2 &=& \sin{\xi_1}
\cos{\xi_2} \,, \\ \sqrt{2}C{_3} &=& \sin{\xi_1} \sin{\xi_2} \,,
\nonumber
\end{eqnarray}
where the constraint is automatically satisfied and the new constants
of integration are $0 \leq \xi_1 \leq \pi$ and $0 \leq \xi_2 < 2\pi$.

Using Eq. (\ref{eq:a}) in Eqs. (\ref{eq:consts})
gives monotonic power law solutions for $v_4$, $v_3$ and $v_2$:
\begin{eqnarray}
\label{eq:sv4}
e^{v_4}&=&e^{{v_{4*}}}|\eta|^{\sin{\xi_1}\sin{\xi_2}/\sqrt{2}} \,,
\nonumber\\ \label{eq:sv3}
e^{v_3}&=&e^{{v_{3*}}}|\eta|^{\sin{\xi_1}\cos{\xi_2}} \,,\\
\label{eq:sv2} e^{v_2}&=&e^{{v_{2*}}}|\eta|^{\sqrt{3}\cos{\xi_1}}
\,, \nonumber
\end{eqnarray}
where $v_{4*}$, $v_{3*}$ and $v_{2*}$ are constants of integration.
Dilaton-moduli-vacuum solutions related by SL(3,R)
transformations~\cite{axion pertbn} are recovered by setting
$\xi_2=0$, while for $\xi_2=0$ and $\xi_1=0$ or $\pi$ we recover dilaton-vacuum
cosmological solutions related by SL(2,R)
transformations~\cite{low energy eff str cos}.

\section{Axion Perturbations}

It is known~\cite{BruGasGio95,GasGio97} that inhomogeneous linear
perturbations in the dilaton and other moduli fields about the
homogeneous FRW solutions given by Eq.~(\ref{eq:a}) have the general
solution $\delta v_i = Z_0 (|k\eta|)$, where $Z_0$ is any linear
combination of Bessel functions of order zero, independent of the
various integration constants that appear in the solutions in
Eq.~(\ref{eq:sv4}). In the pre-big-bang scenario this solution
inevitably leads to a cosmological spectrum of vacuum fluctuations
steeply tilted towards small scales, with essentially no perturbations
on super-horizon scales~\cite{BruGasGio95,LidWanCop99}.  If the pre
big bang scenario is to produce any observable perturbations on large
scales it must be through vacuum fluctuations in the axion
fields~\cite{vac fluctn}
which are non-minimally coupled to the dilaton-moduli fields in the
four-dimensional Einstein frame and hence sensitive to the integration
constants that parameterise their evolution.

Hence we will investigate inhomogeneous axion field perturbations
about the dilaton-moduli vacuum solutions.
We will calculate the field equations for these linear
perturbations including axion fields by constructing the effective
action to second-order in the perturbations\footnote {Because
$\sigma_i'=0$ in the background solution there is no metric
back-reaction to lowest-order and the perturbations are
automatically gauge-invariant~\cite{LidWanCop99}.}. This is
constructed iteratively using Eq.~(\ref{slnvac}) as
\begin{eqnarray}
{\mathrm{Tr}}\left({\nabla}U_{n+1} {\nabla}{U_{n+1}}^{-1} \right)
& = & {\mathrm{Tr}}\left({\nabla}U_n{\nabla}{U_n}^{-1}\right)
\nonumber\\&& +(-n^2-n)\left({\nabla}v_{n+1}\right)^{2}
\nonumber\\ && \hspace{-36pt} -2{\mathrm{Tr}}\left(
e^{(n+1)v_{n+1}} {\nabla}{\sigma}_n^{T} U{_n}^{(0)}
{\nabla}{\sigma}_n \right) \,,
\end{eqnarray}
where we can use the vacuum solution, $U_n^{(0)}$, in the last term of
the trace equation, in order to calculate the action to second-order
in the axion fields.

Expressions can be constructed from the initial SL(2,R) matrix
Eq. (\ref{eq:sl2matrix}) and we can therefore write the trace of
${\nabla}U_3{\nabla}{U_3}^{-1}$ for SL(3,R) to second order in the
axion perturbations as
\begin{eqnarray}
{\mathrm{Tr}}\left({\nabla}U_3 {\nabla}U{_3}^{-1} \right) &=&
{\mathrm{Tr}}\left({\nabla}U_2{\nabla}U{_2}^{-1}\right)
-6\left({\nabla}v_2\right)^{2} \nonumber\\ && - 2{\mathrm{Tr}}
\left(e^{3v_2}{\nabla}{\sigma}^{T}U{_2}^{(0)}{\nabla}{\sigma}\right)
\,.
\end{eqnarray}
Rewriting Eq. (\ref{eq:sl2matrix}) as
\begin{equation}
U_2=
\left( \begin{array}{ll}
e^{v_2} & \sigma_1 e^{v_2}\\
\sigma_1 e^{v_2} & e^{-v_2}+\sigma_1^{2}e^{v_2}\\
\end{array}  \right)
\end{equation}
and then adding on the second order axion term with the vacuum
case given in Eq.~(\ref{slnvac}) in the above expression we
obtain~\cite{axion pertbn}
\begin{eqnarray}
{\mathrm{Tr}}\left({\nabla}U_3{\nabla}{U_3}^{-1}\right)
&=&
-6({\nabla}v_3)^2-2({\nabla}v_2)^2
-2e^{2v_2}({\nabla}\sigma_1)^2 \nonumber\\
&&
\hspace{-24pt}
-2e^{3v_3+v_2}({\nabla}\sigma_2)^2
-2e^{3v_3-v_2}({\nabla}\sigma_3)^2 \,,
\end{eqnarray}
while for SL(4,R) we have
\begin{eqnarray}
{\mathrm{Tr}}\left({\nabla}U_4{\nabla}{U_4}^{-1}\right)
& = & \nonumber\\
&& \hspace{-80pt}
-12({\nabla}v_4)^2
-6({\nabla}v_3)^2-2({\nabla}v_2)^2
-2e^{2v_2}({\nabla}\sigma_1)^2 \nonumber\\
& &  \hspace{-80pt}
-2e^{3v_3+v_2}({\nabla}\sigma_2)^2
-2e^{3v_3-v_2}({\nabla}\sigma_3)^2
-2e^{4v_4+v_3+v_2}({\nabla}\sigma_4)^2\nonumber\\
& &  \hspace{-80pt}
-2e^{4v_4+v_3-v_2}({\nabla}\sigma_5)^2
-2e^{4v_4-2v_3}({\nabla}\sigma_6)^2 \,.
\end{eqnarray}
In the Einstein frame the action becomes
\begin{eqnarray}
S&=&\frac{1}{2\kappa^2}\,\int {\mathrm{d^3}}x\int
{\mathrm{d}}\eta\ \ \Big[-6{a'}^2
-3a^2{{v_4}'}^2-\frac{3}{2}\,a^2{{v_3}'}^2\nonumber\\ & &
-\frac{1}{2}\,a^2{{v_2}'}^2 -\frac{1}{2}
\,a^2e^{2v_2}{{\sigma_1}'}^2-\frac{1}{2}
\,a^2e^{3v_3+v_2}{{\sigma_2}'}^2\nonumber\\ & & -\frac{1}{2}
\,a^2e^{3v_3-v_2}{{\sigma_3}'}^2
-\frac{1}{2}\,a^2e^{4v_4+v_3+v_2}{{\sigma_4}'}^2\nonumber\\ & &
-\frac{1}{2} \,a^2e^{4v_4+v_3-v_2}{{\sigma_5}'}^2 -\frac{1}{2}
\,a^2e^{4v_4-2v_3}{{\sigma_6}'}^2\Big] \,.
\end{eqnarray}
We can now use this to derive the field equations for inhomogeneous
axion perturbations evolving in the homogeneous vacuum background
solutions
\begin{eqnarray}
\label{eq:sigma1p}{\delta\sigma_1}''+
\left(2\frac{a'}{a}+2{v_2}'\right){\delta\sigma_1}'+k^2\delta\sigma_1&=&0
\,,\\ \label{eq:sigma2p}{\delta\sigma_2}''+
\left(2\frac{a'}{a}+3{v_3}'+{v_2}'\right){\delta\sigma_2}'+k^2\delta\sigma_2
&=&0 \,,\\ \label{eq:sigma3p}{\delta\sigma_3}''+
\left(2\frac{a'}{a}+3{v_3}'-{v_2}'\right){\delta\sigma_3}'+k^2\delta\sigma_3
&=&0 \,,\\ \label{eq:sigma4p}{\delta\sigma_4}''+
\left(2\frac{a'}{a}+4{v_4}'+{v_3}'+{v_2}'\right){\delta\sigma_4}'+k^2\delta\sigma_4
&=&0 \,,\\ \label{eq:sigma5p}{\delta\sigma_5}''+
\left(2\frac{a'}{a}+4{v_4}'+{v_3}'-{v_2}'\right)
{\delta\sigma_5}'+k^2\delta\sigma_5 &=&0 \,,\\
\label{eq:sigma6p}{\delta\sigma_6}''+
\left(2\frac{a'}{a}+4{v_4}'-2{v_3}'\right){\delta\sigma_6}'
+k^2\delta\sigma_6&=&0 \,,
\end{eqnarray}
where $k$ is the comoving wavenumber.
Note that because the axion field is zero in the vacuum background
solution, their back-reaction upon the moduli fields and the
four-dimensional spacetime metric vanishes to first-order and the
perturbations are independent of the spacetime
gauge~\cite{LidWanCop99}.

The field equations (\ref{eq:sigma1p}) to (\ref{eq:sigma1p}) can be
written in the standard form for a free scalar field evolving in a FRW
metric
\begin{equation}
\label{eq:pertbn}
{\delta\sigma_i}''
+2\frac{{\bar{a}_i'}}{\bar{a}_i}{\delta\sigma_i}'+k^2\delta\sigma_i=0
\end{equation}
with $i=1 \ldots 6$, where we introduce the conformally rescaled scale
factor, ${\bar{a}_i}$, in the corresponding ``axion frame'' \cite{vac
fluctn,LidWanCop99}
\begin{eqnarray}
\label{eq:bara1} \bar{a}_1 &=& e^{v_2} a \,,\\
\label{eq:bara2} \bar{a}_2 &=& e^{(3v_3+v_2)/2} a \,,\\
\label{eq:bara3} \bar{a}_3 &=& e^{(3v_3-v_2)/2} a \,,\\
\label{eq:bara4} \bar{a}_4 &=& e^{2v_4+(v_3+v_2)/2} a \,,\\
\label{eq:bara5} \bar{a}_5 &=& e^{2v_4+(v_3-v_2)/2} a \,,\\
\label{eq:bara6} \bar{a}_6 &=& e^{2v_4-v_3} a \,.
\end{eqnarray}
Substituting the solutions from Eqs.~(\ref{eq:a}) and
(\ref{eq:sv3}) into Eqs.~(\ref{eq:bara1})--(\ref{eq:bara6}) we
obtain
\begin{equation}
\label{eq:bara}
\bar{a}_i \propto |\eta|^{p_i} \,,
\end{equation}
where
\begin{eqnarray}
\label{eq:pi} p_1 &=& \sqrt{3} \cos{\xi_1} \,, \nonumber\\ p_2 &=&
\sqrt{3} \left( {\sqrt{3}\over2}\sin\xi_1\cos\xi_2 +
{1\over2}\cos\xi_1 \right)
\,, \nonumber \\
 p_3 &=&
\sqrt{3} \left( {\sqrt{3}\over2}\sin\xi_1\cos\xi_2 -
{1\over2}\cos\xi_1 \right)
\,, \nonumber \\
 p_4 &=&
\sqrt{3} \left[ {\sqrt{3}\over2}\sin\xi_1 \left(
{2\sqrt{2}\over3}\sin\xi_2 + {1\over3}\cos\xi_2 \right) +
{1\over2}\cos\xi_1 \right]
\,, \nonumber\\
 p_5 &=&
\sqrt{3} \left[ {\sqrt{3}\over2}\sin\xi_1 \left(
{2\sqrt{2}\over3}\sin\xi_2 + {1\over3}\cos\xi_2 \right) -
{1\over2}\cos\xi_1 \right]
\,, \nonumber \\
p_6 &=& \sqrt{3} \sin{\xi_1} \left(
{\sqrt{2}\over\sqrt{3}}\sin{\xi_2}
 - {1\over\sqrt{3}}\cos{\xi_2} \right) \,.
\end{eqnarray}

Equation~(\ref{eq:pertbn}) has the standard form for perturbations of
a free scalar field evolving in an FRW
 cosmology with scale factor
$\bar{a}_i$. Thus we can define the canonically normalised variables
\cite{birrel,LidWanCop99}
\begin{equation}
\label{eq:variable}
u_i=\frac{1}{\sqrt{2}\kappa}\,\bar{a}_i\delta\sigma_i \,,
\end{equation}
which enables us to rearrange Eq.~(\ref{eq:pertbn}) in the form of a
simple harmonic oscillator with time-dependent mass
\begin{equation}
\label{eq:bessel}{u_i}''
+\left(k^2+\frac{({-\bar{a}_i}'')}{\bar{a}_i}\right)u_i=0 \,.
\end{equation}
For a power-law expansion given by Eq.~(\ref{eq:bara}) this
corresponds to Bessel's equation
\begin{equation}
\label{eq:bessel2}{u_i}''
+\left(k^2+\frac{(1/4-p{_i}^2)}{{|\eta|}^2}\right)u_i=0 \,.
\end{equation}
Equation~(\ref{eq:bessel2}) has a standard solution
\begin{equation}
u_i=|k \eta|^{1/2} Z_{p_i}\left(|k \eta|\right) \,,
\end{equation}
where $Z_{p_i}$ is any linear combination of Bessel (or Hankel)
functions and for each axion field the order of the Bessel function is
$p_i$ given in Eq.~(\ref{eq:pi}).

\section{Pre-Big-Bang Perturbation Spectra}

We will write the solutions of the Bessel equation in terms of
Hankel functions so that a general solution of Eq. 
(\ref{eq:bessel2}) is given by
\begin{equation}
u_i=|k\eta|^{1/2}
\left[u_+{H_{|p_i|}}^{(1)}\left(|k\eta|\right)
+u_-{H_{|p_i|}}^{(2)}\left(|k\eta|\right)\right] \,.
\end{equation}

For wavelengths much smaller than the horizon scale ($|k\eta|\gg1$)
the equation of motion Eq. (\ref{eq:bessel}) reduces to that for a
free-scalar field $u_i$ in flat Minkowski spacetime with a
well-defined vacuum state. Allowing only
positive frequency modes in a flat-spacetime vacuum state requires
that
\begin{equation}
u_i \to \frac{e^{-ik \eta}}{\sqrt{2k}} \,.
\end{equation}
The classic horizon problem of the standard hot big bang is that all
modes start outside the horizon at the big bang ($k\eta\to0$) and so
there is no reason to expect modes to start in this vacuum state.

However in the pre-big-bang scenario all modes start within the
horizon in the infinite past as $\eta\to-\infty$. As $\eta\to0_-$ on
the $(+)$ branch modes leave the horizon, giving a well-defined
spectrum of super-horizon perturbations in all fields, even though
there is no inflation in the conventional sense ($\ddot{a}>0$) in the
Einstein frame.

Allowing only positive frequency modes in a flat-spacetime vacuum
state at early times (as $k \eta \to - \infty$) i.e. large $-\eta$
%
%
yields
\begin{equation}
u_+ = \frac{\sqrt{\pi}}{2\sqrt{k}} \; e^{i(2|p_i|+1)\pi/4} \, , \quad
u_- = 0 \,.
\end{equation}
Therefore using Eq.  (\ref{eq:variable}) we can write
\begin{equation}
\label{eq:deltasigma1}\delta\sigma_i =\kappa
\sqrt{\frac{\pi}{2k}}\; e^{i(2|p_i|+1)\pi/4}\;
\frac{\sqrt{-k\eta}}{\bar{a}_i}\;
{H_{|p_i|}}^{(1)}\left(-k\eta\right) \,.
\end{equation}
At late times on super-horizon scales ($|k\eta|\ll1$) we have
%
%
%
%
%
%
\begin{equation}
\label{eq:deltasigma2}\delta\sigma_i=\pm i \kappa
\sqrt{\frac{\pi}{k}}\; e^{i(2|p_i|+1)\pi/4}\;
\left(\frac{-\Gamma(|p_i|)}{\pi\bar{a}_i}\right) \left(\frac{2}{-k
\eta}\right)^{|p_i|-(1/2)} \,.
\end{equation}
%

The power spectrum for these axion perturbations is denoted by
\begin{equation}
{\mathcal{P}}_{\delta\sigma_i} \equiv
\frac{k^3}{2\pi^2}|\delta\sigma_i|^2
\end{equation}
which represents the dispersion $\langle\delta\sigma_i^2\rangle$
due to fluctuations on comoving scales $\sim k^{-1}$~\cite{LidLyt93}.
It can be calculated from Eq.~(\ref{eq:deltasigma2})
%
%
to be
\begin{equation}
\label{eq:pspectra}
{\mathcal{P}}_{\delta\sigma_i}
=2\kappa^2
\left(\frac{C(|p_i|)}{2\pi}\right)^2\frac{k^2}{{\bar{a}_i}^2}(-k\eta)^{1-2|p_i|}
\,,
\end{equation}
where the coefficient term
\begin{equation}
\label{eq:Cpi}C(|p_i|)=\frac{2^{|p_i|}\Gamma(|p_i|)}{2^{3/2}\Gamma(3/2)} \,,
\end{equation}
approaches unity for $|p_i|=3/2$.

The spectral tilt of the perturbation spectra is defined by
\begin{equation}
\Delta n_i \equiv \frac{\mathrm{d}\ln
{\mathcal{P}}_{\delta\sigma_i}}{{\mathrm{d}} \ln k} \,.
\end{equation}
It follows from Eq.  (\ref{eq:pspectra}) that the spectral
tilt for each of the axion fields is constant and takes the values
\begin{equation}
\label{eq:indices} \Delta n_i=3-2|p_i| \,,
\end{equation}
where the $p_i$ are given in Eq.~(\ref{eq:pi}) in terms of the two
integration constants $\xi_1$ and $\xi_2$ which parameterise the
dilaton-moduli vacuum background solutions.

\begin{figure}[t]
\centering
\leavevmode\epsfysize=5cm \epsfbox{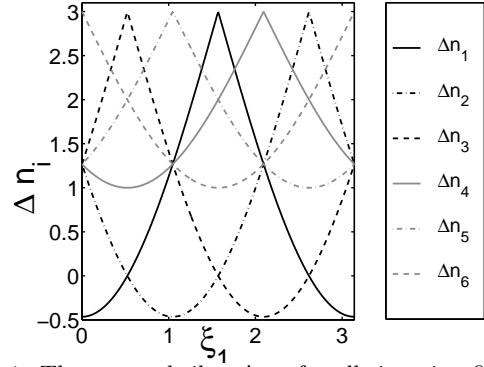}\\
\caption[Figure1]{The spectral tilts, $\Delta n_i$, for all six
axion fields as a function of $\xi_1$ when $\xi_2=0$.}
\label{fig1}
\end{figure}

\begin{figure}[t]
\centering
\leavevmode\epsfysize=5cm \epsfbox{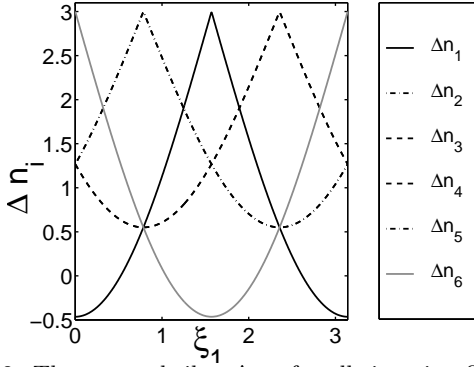}\\
\caption[Figure2]{The spectral tilts, $\Delta n_i$, for all six
axion fields as a function of $\xi_1$ when $\tan\xi_2=-\sqrt{2}$.}
\label{fig2}
\end{figure}

{}From Eqs. ~(\ref{eq:indices}) and~(\ref{eq:pi}) it is then
possible to find the spectral tilts as functions of the
integration constants $\xi_1$ and $\xi_2$. (See Figs.~\ref{fig1}
and~\ref{fig2}.) The maximum absolute value for any $p_i$ is
$\sqrt{3}$ and thus the minimum value of the spectral tilt for any
axion field is $\Delta n = -2\sqrt{3}+3=-0.46$ as found previously
with SL(2,R) \cite{vac fluctn} or SL(3,R) \cite{axion pertbn}
symmetry groups. For any axion field the allowed range for the
spectral tilt is
\begin{equation}
-2\sqrt{3}+3 \leq \Delta n_i \leq 3 \,.
\end{equation}


\section{Discussion}

The amplitude of the axion power spectra at the end of the pre-big
bang phase when the comoving horizon scale is given by $\eta_s=-1/k_s$
can be given, from Eqs.~(\ref{eq:pspectra}) and~(\ref{eq:indices}), as
\begin{equation}
\label{eq:amplitude} {\mathcal P}_{\delta \sigma_i} |_s = 2
\kappa^2 \left(\frac{C(|p_i|)}{p_i + 1/2}\right)^2
\left(\frac{\bar{H_i}}{2 \pi}\right)_s^2
\left(\frac{k}{k_s}\right)^{\Delta n_i} \,,
\end{equation}
where the expression for $C(|p_i|)$ is given in Eq. 
(\ref{eq:Cpi}) and the Hubble rate in the axion
frame is given by
\begin{equation}
\bar{H_i}^2 = \frac{(p_i + 1/2)^2}{\bar{a}_i^2 \eta^2} \,.
\end{equation}

The Hubble rate in the axion frame can also be related to $H$, the
expansion rate in the Einstein frame, via
\begin{equation}
\bar{H_i} = \frac{2 (p_i + 1/2)}{\Omega_i} H \,.
\end{equation}
Assuming that modes remain frozen-in on large scales $(|k \eta| \ll
1)$ during the uncertain transition from pre to post big bang
phase~\cite{BruGasVen98,BruHad98,Buonanno}, then the energy density
contributed by massless axion perturbations in the Einstein frame can
be estimated as
\begin{eqnarray}
\rho_i &\approx&
 \Omega{_i}^2 \frac{k^2}{a^2}
\frac{{\mathcal P}_{\delta \sigma_i} |_s}{2 \kappa^2}\\
%
%
&\approx& \frac{k^2}{a^2} C^2(|p_i|)
\left(\frac{H_s}{2\pi}\right)^2 \left(\frac{k}{k_s}\right)^{\Delta n_i} \,.
\end{eqnarray}

Although massless axion fields never
come to dominate the total energy density in the universe, they do
contribute a perturbation to the density when a given scale re-enters
the cosmological horizon
\begin{equation}
\label{eq:deltarho}
\frac{\delta\rho}{\rho} =
\frac{\rho_i}{\rho_{\rm total}} = {C^2(|p_i|)\over3}
\left(\frac{\kappa H_s}{2\pi}\right)_s^2
\left(\frac{k}{k_s}\right)^{\Delta n_i} \,,
\end{equation}
where the total energy density during the radiation or matter
dominated era is given by $\rho_{\rm total}=3H^2/\kappa^2$.  An
analysis of the effect of this perturbation to the overall density
during the radiation and matter dominated eras, and hence to the
spectrum of anisotropies in the cosmic microwave background, is
calculated in Refs.~\cite{durrer98,durrer99}.  Note that all the
different axion fields have the same amplitude for modes crossing the
Hubble scale at the start of the post big bang era, $k\approx k_s$,
set by the ratio of the maximum Hubble rate at the end of the pre big
bang to the Planck scale $\kappa^2H_s^2$.  In the simplest scenario of
a sudden transition to the post big bang phase where the dilaton and
moduli are fixed, this gives $\kappa^2H_s^2\approx e^\phi\approx
10^{-2}$ yielding the expression given in Eq.~(\ref{one}). On larger
scales the amplitudes of the different axion fields is then fixed
entirely by the different spectral tilts, with the field with the
minimum tilt yielding the largest density perturbation. If $\Delta
n_i<0$ for any of the axion fields then the density perturbation given
by Eq.~(\ref{eq:deltarho}) diverges on large scales.

A key question then is whether it is possible to maintain blue
spectral indices ($\Delta n_i>0$) for {\em all} the axion fields for
any values of $\xi_1$ and $\xi_2$.  Figure~\ref{fig1} shows the
spectral indices for solutions restricted to the SL(3,R)
dilaton-moduli-vacuum solutions where $\xi_2=0$, in which case it has
previously been shown~\cite{axion pertbn} that there are no values of
$\xi_1$ for which $\Delta n_i>0$ for all $i$.

\begin{figure}
\centering
\leavevmode\epsfysize=5cm \epsfbox{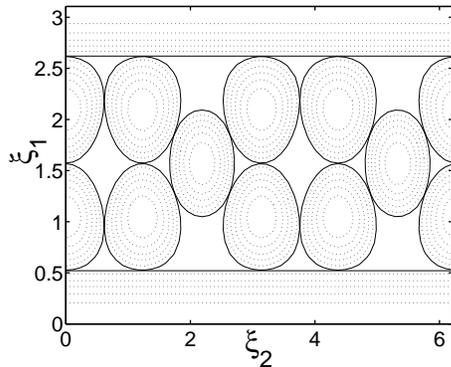}\\
\caption[figure3]{Contour plot of the minimum value of all six
spectral indices, $\Delta n_*$, as a function of integration
constants $\xi_1$ and $\xi_2$. The solid line is the contour
$\Delta n_*=0$, and dotted lines represent contours for $\Delta
n_*<0$, where at least one of the axion spectral indices is
negative.} \label{fig3}
\end{figure}

We define $\Delta n_* \equiv \mathrm{min} (\Delta n_1,\Delta
n_2,\ldots,\Delta n_6)$ to be the smallest of all the spectral indices
for a given value of $(\xi_1,\xi_2)$.
Figure~\ref{fig3} shows that there is a finite region of parameter
space for which $\Delta n_*>0$. We find the maximum value of
$\Delta n_*$ occurs for $\xi_1=\pi/4$ or $3\pi/4$ and
$\tan\xi_2=-\sqrt{2}$ (shown in Fig.~\ref{fig2} where we obtain
\mbox{$\Delta n_*=3-\sqrt{6}=0.55$}).  Table 1 shows the previously
calculated maximum values for $\Delta n_*$ for other SL($n$,R)
groups.  The maximum allowed spectral tilt gets progressively
larger (bluer) as the symmetry group gets larger. Since the larger
groups always contain the previous groups we see that it is indeed
possible for all the axion fields to have blue spectral indices in
SL($n$,R) non-linear sigma models where $n\geq4$.
\begin{table}[h]
\begin{tabular}{|c|c|}
Symmetry Group & maximum $\Delta n_*$ \hspace{2cm} \\
\hline
SL(2,R) & -0.46 \hspace{2cm} \\
SL(3,R) & 0  \hspace{2cm} \\
SL(4,R) & 0.55  \hspace{2cm} \\
\end{tabular}
\caption{Summary of maximum allowed value of $\Delta n_*$ for
successive SL($n$,R) groups.}
\end{table}
On the other hand requiring all the spectral indices to be positive
represents a restriction on the allowed initial conditions.

We have focused our discussion on the cosmological perturbations
induced by an effectively massless axion field proposed in
Ref.~\cite{durrer98}, in which case the primordial axion
perturbation spectra can be directly related to seed density
perturbations. 
If the axions become non-relativistic during the radiation-dominated
era then the overall amplitude of perturbations is increased by a
factor $(m/H_{\rm eq})^{1/2}$~\cite{durrer98,BruHad98,GasVen99}, where
the Hubble rate at matter-radiation equality is $H_{\rm
eq}\sim10^{-27}$eV.
If any of the primordial axion spectra have a negative spectral tilt,
$\Delta n_i<0$ (as occurs in a large regime of parameter space shown
in Fig.~\ref{fig3}) then the only way to suppress the large-scale
density perturbation appears to be for the axion to acquire a periodic
potential where large field fluctuations may have a small effect on
the energy density~\cite{KofLin87}.  In such a scenario, the massive
axion could be a novel form of dark matter and lead to a very
different model for large-scale structure formation\cite{BruHad98b}.

Finally, we note that the presence of non-trivial background axion
fields in the general solutions of the full SL(2,R) and SL(3,R)
symmetric axion-dilaton-moduli cosmologies restricts the allowed
asymptotic vacuum state~\cite{low energy eff str cos,IIB,Alex}. The
effect of the background axion fields upon the perturbation spectra
has only been calculated for a single axion field where the
perturbation spectra remain invariant under SL(2,R) transformations of
the background solutions~\cite{vac fluctn,dual}. The cosmological
perturbation spectra in axion-dilaton-moduli solutions with more
degrees of freedom remain to be investigated.

\section*{Acknowledgements}

The authors are grateful to Carlo Ungarelli for helpful comments.
H.B. is supported by the EPSRC and D.W. is supported by the Royal Society.

\end{document}